\documentclass[%
 reprint,
 amsmath,amssymb,
 aps,
]{revtex4-2}

\usepackage{graphicx}
\usepackage{dcolumn}
\usepackage{bm}
\usepackage{cancel}

\usepackage[usenames, dvipsnames]{xcolor}
\usepackage{tikz}
\usetikzlibrary{shapes}
\usepackage[co mpat=1.1.0]{tikz-feynman}

\begin{document}


\title{Hidden Supersymmetric Dark Sectors}

\author{Giacomo Cacciapaglia}
\email{g.cacciapaglia@ip2i.in2p3.fr}
\affiliation{IP2I, Universit{\'e} de Lyon, UCBL, UMR 5822 CNRS/IN2P3, 4 rue Enrico Fermi, 69622 Villeurbanne Cedex, France}

\author{Aldo Deandrea}%
\email{deandrea@ip2i.in2p3.fr }
\affiliation{IP2I, Universit{\'e} de Lyon, UCBL, UMR 5822 CNRS/IN2P3, 4 rue Enrico Fermi, 69622 Villeurbanne Cedex, France}
\affiliation{Department of Physics, University of Johannesburg, PO Box 524, Auckland Park 2006, South Africa}

\author{Wanda Isnard}
\email{wanda.isnard@ens-lyon.fr}

\affiliation{ Department of Physics, \'{E}cole Normale Supérieure de Lyon, \\
 46 Allée d'Italie, 69364 Lyon Cedex 07, France}

\begin{abstract}
The attractive feature of supersymmetry is predictive power, due to the large number of calculable properties and to coupling non-renormalisation. This power can be fully expressed in hidden sectors where supersymmetry may be exact, as these sectors are secluded from the visible one where instead supersymmetry must be broken. This suggests a new paradigm for supersymmetric dark sectors, where supersymmetry is exact at the dark matter scale, implying that many properties of hidden supersymmetric dark sectors can be fully computed. As a proof of concept we discuss a concrete example based on $\mathcal{N}=1$ super Yang-Mills.
\end{abstract}

\maketitle


Supersymmetry \cite{Gervais:1971ji,Golfand:1971iw,Wess:1974tw} is arguably the most natural extension of Minkowsky space-time symmetries, beyond the Poincar{\'e} algebra \cite{Haag:1974qh}. It links bosonic and fermionic particles, hence relating integer and semi-integer spins \cite{Miyazawa:1966mfa}.
Furthermore, supersymmetric field theories enjoy special properties that make for an exciting theoretical playground. For instance, the non-renormalisation of a set of operators \cite{Wess:1974tw,Weinberg:1998uv} renders supersymmetry a prime candidate to explain the quantum stability of scalar masses \cite{Seiberg:1993vc}, in particular for the Higgs boson. As a second example, theories with extended supersymmetries can be solved exactly without the need for a perturbative expansions \cite{Seiberg:1988ur}.
This idyllic theoretical situation contrasts with the physical real world, where the Standard Model of particle physics (SM) lacks any supersymmetric feature, {\it  in primis} a mirror fermion-boson symmetric particle content. Henceforth, if supersymmetry is realised, it must be at energy scales well beyond the ones already probed at accelerators, like the Large Hadron Collider (LHC) at CERN, or in secluded sectors, at most feebly connected to the SM.
However, there are phenomena that are not captured by the SM, specifically the presence of Dark Matter (DM), which constitutes 75\% of the matter budget of the present-day Universe \cite{Planck:2018nkj}. The only DM direct evidence involves its gravitational interactions, which have been observed at a variety of scales: from galaxies, via the rotational velocity of luminous stars and gas, to gravitational lensing of galaxy clusters, to ultimately the global properties of the observable Universe, revealed via the Cosmic Microwave Background. Many models propose a particle candidate to represent DM, including particles predicted by (broken) supersymmetry \cite{Roszkowski:2017nbc,Feng:2022rxt} and hidden sectors \cite{Lagouri:2022oxv}. See Refs~\cite{Bauer:2017qwy,Lin:2019uvt,Arbey:2021gdg} for comprehensive reviews on DM models and properties. For a long time, the Minimal Supersymmmetric SM (MSSM) has offered a standard DM candidate in the form of the lightest supersymmetric particle (LSP), which is usually a neutralino or gravitino. Its stability, however, is not guaranteed directly by supersymmetry as it requires an imposed matter symmetry (R--parity), which is also necessary to forbid proton decay \cite{Barbier:2004ez}. As the DM candidate is the lightest partner of the SM states, its mass and nature in the LSP picture is strongly related to the breaking of supersymmetry.

In this letter we propose a new paradigm of supersymmetric DM (sDM), alternative to the LSP scenario, as it is characterised by an approximate supersymmetry at the scale of the DM mass. This can be realised by adding a dark sector to the MSSM: either the new supersymmetric states are heavier than the breaking scale of the MSSM, or they belong to a hidden sector, feebly coupled to the MSSM fields.  The latter has the advantage of benefiting from supersymmetric properties, primarily its calculability. Models of hidden supersymmetric sectors have been considered in the literature under various motivations \cite{Arnold:2012wm,Heikinheimo:2013xua,Dery:2019jwf,Csaki:2022xmu}. Henceforth, the main model building ingredient is a hierarchy between the supersymmetry breaking scales in the visible sector, needed to be above a few TeV due to collider bounds, and in the Dark sector, where supersymmetry is assumed to be valid at the DM mass scale. Also, the properties of the hidden supersymmetric sector are disjointed from phenomenological requirements related to the SM physics, opening a new set of simple and attractive possibilities. 

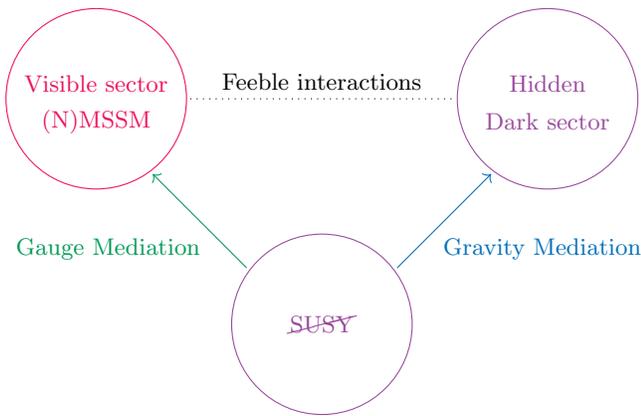
\begin{figure}[htbp]
\centering
	\begin{tikzpicture}
		\draw [color=OrangeRed](-3,0) circle (1.2);
		\draw [color=Purple] (3,0) circle (1.2);
		\draw [color=Purple] (0,-3) circle (1.2);

		\draw (-3,0.2) node[OrangeRed]{Visible sector};
		\draw (-3,-0.3) node[OrangeRed]{(N)MSSM};
		\draw (3,0.2) node[Purple]{Hidden};
        \draw (3,-0.3) node[Purple]{Dark sector};
		\draw (0,-3) node[Purple]{\cancel{SUSY}};

		\draw[->,color=ForestGreen] (-1,-2.25) -- (-2.25,-1);
		\draw[->,color=NavyBlue] (1,-2.25) -- (2.25,-1);couleur
		\draw (-1.5,-2) node[left,ForestGreen]{Gauge Mediation};
		\draw (1.5,-2) node[right,NavyBlue]{Gravity Mediation};

		\draw [dotted] (-1.75,0) -- (1.75,0);
		\draw node[above]{Feeble interactions};
	\end{tikzpicture}
	\caption{\label{Model} Illustration of the different sectors of the model with a hidden sDM. Purple circles are hidden sectors while the red circle is the visible sector of the theory. The source of supersymmetry breaking (\cancel{SUSY}) is transmitted to the visible and the hidden sectors through gauge and gravity mediation, respectively. There are also feeble interactions between the visible and the hidden sectors.}
\end{figure}

A schematic realisation of this sDM scenario is illustrated in Fig.~\ref{Model}: both the visible and hidden dark sectors consist of supersymmetric theories, connected to each other via feeble interactions (a similar setup was considered in \cite{Dery:2019jwf}). Those can consist of a small supersymmetric coupling \cite{Csaki:2022xmu} or interactions due to a heavy mediator. As such, the dark sector can be populated via the freeze-in mechanism \cite{Hall:2009bx} to generate the correct DM relic density. Both sectors are connected to the same supersymmetry breaking source (\cancel{SUSY}), however the breaking is mediated by different mechanisms. To generate a clear hierarchy between the \cancel{SUSY} scales in the two sectors, we choose gravity mediation \cite{Chamseddine:1982jx,Barbieri:1982eh,Hall:1983iz} for the hidden sector and gauge mediation \cite{Dine:1981gu,Nappi:1982hm,Alvarez-Gaume:1981abe} for the visible sector. Hence, the model set-up proposed in Fig.~\ref{Model} draws a connection between the Planck scale $M_P=10^{19}$~GeV and the supersymmetry breaking scales in the visible sector, $m_{\rm VS}$, and in the hidden sector, $m_{\rm HS}$. We recall that the latter must be smaller than the DM mass, $m_{\rm HS} \ll m_{\rm DM}$. Assuming, as an illustration, that the \cancel{SUSY} scale is generated via $F$-term breaking, the scales are related as follows:
\begin{eqnarray}
    m_{\rm{HS}} \sim \frac{\langle F_X \rangle}{M_P}\,, \quad
	m_{\rm{VS}} \sim \frac{g_G^2}{16 \pi^2} \frac{\langle F_X \rangle}{M_G}\,,
\end{eqnarray}
where $g_G$ is the gauge coupling of the heavy gauge mediators of mass $M_G$, which needs to sit well below the Planck scale for consistency. Structure formation typically requires the DM mass to be above the $100$~keV scale \cite{Boehm:2003xr,Hooper:2007tu}. 
Hence, imposing $m_{\rm HS} < 100$~keV and $m_{\rm VS} = 10$~TeV, yields the following estimates:
\begin{eqnarray}
\langle F_X \rangle \lesssim 10^{15} \; \text{GeV}^2\,, \quad 
M_G \lesssim 10^9 \; \text{GeV}\,,
\end{eqnarray}
for $g_G \sim 1$. We shall retain them as typical orders of magnitude for the model building. Note that $m_{\rm VS}$ can also contribute to $m_{\rm HS}$ via the interactions between these two sectors, hence validating the assumption that they must be feeble. This scenario potentially suffers from the cosmological gravitino problem \cite{Weinberg:1982zq}. Gravitinos are the spin-3/2 super-partners of gravitons, and they are produced in the early Universe via gravitational interactions. Their presence at late times can efficiently suppress structure formation. One possible solution is to lower their mass at the eV scale \cite{Moroi:1993mb}, with most recent bounds reading $m_{3/2} \lesssim 4.7$~eV \cite{Osato:2016ixc}. As $m_{3/2} \sim m_{\rm HS}$, this mass limit fits well within our scenario if 
\begin{equation}
\langle F_X \rangle \lesssim 10^{10}~\text{GeV}^2\,.
\end{equation}
The sDM candidate, which is the lightest state in the hidden supersymmetric sector, could decay into gravitinos, hence repopulating them in the late universe. There are two kinds of decays: channels involving the visible sector, doubly suppressed by the Planck mass and the feeble interactions; and decays among components of the same multiplet in the hidden sector, suppressed by the small mass splitting and on-threshold masses (as $\Delta m \sim m_{3/2} \sim m_{\rm HS}$). This deems our sDM scenario generally free from the cosmological gravitino problem.

\vspace{0.3cm}

\begin{figure}[htbp]
\centering
\includegraphics[height=5.2cm]{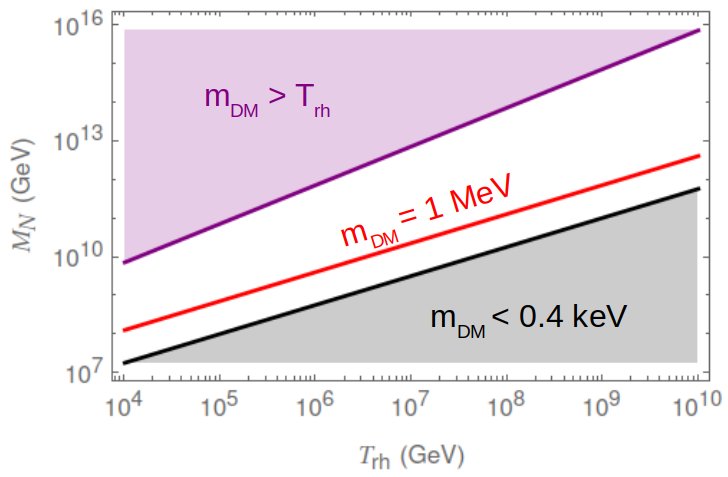}
\caption{\label{fig:relic} Suitable values of $T_{\rm{rh}}$ and $M_N$ for obtaining the observed DM relic density via ultraviolet freeze-in in the SYM model for $N_c=3$. For different values of $N_c$, $m_{\rm DM}$ is re-scaled to keep $N_c\ m_{\rm DM}$ fixed. The preferred region lies between the red line and the lower edge of the lilac region.}
\end{figure}

As a concrete, simple and calculable example of the hidden dark sector, we consider a $\mathcal{N}=1$ supersymmetric Yang-Mills (SYM) theory \cite{Wess:1974tw}, based on the gauge symmetry SU($N_c$). This class of theories possesses classical scale-invariance, which is however broken at quantum level (except for $\mathcal{N}=4$ SYM, which is fully integrable). Hence, like Quantum Chromodynamics, the theory is expected to confine at low energies and generate a dynamical mass scale $\Lambda$. This scale, which is fully supersymmetric, controls the mass of the lightest states in the low energy theory. As no SM states are charged under the hidden SU($N_c$) gauge symmetry, heavy mediators need to be introduced to couple the hidden sector to the visible one. The simplest possibility is to add a set of matter superfields \cite{Ferrara:1974ac} $Q$ and $\tilde{Q}$ in conjugate representations of SU($N_c$), which couple to a gauge-singlet chiral superfield $\hat{N}$ via the superpotential
\begin{equation}
	\int d^2 \theta\ \left[ \lambda_{NQQ}\  \hat{N} Q \tilde{Q} + M_Q\ Q \tilde{Q} \right] + \mbox{h.c.}\,
\end{equation}
Integrating out the massive multiplet, we obtain an effective coupling between the SYM gauge superfields $W^\alpha$ and the singlet superfield
\begin{multline}
    \mathcal{L} \supset \frac{1}{32 \pi} {\rm{Im}} \left( \tau \int d^2 \theta\  {\rm{Tr}}\ W^{\alpha} W_{\alpha} \right) \times \\
    \left( 1 + \frac{1}{M_N^2} \hat{N} \hat{N}^{\dagger} + ...  \right)\,,
    \label{SYMeff}
\end{multline}
where
\begin{equation}
    \frac{1}{M_N^2} \sim \frac{\lambda_{NQQ}^2}{16 \pi^2\ M_Q^2}\,,
\end{equation}
and the dots represent higher order interactions. The effective scale $M_N$ determines the production rate of the dark sector states from the thermal bath of the visible sector, to which the superfield $\hat{N}$ belongs. To simplify the numerical results, we assume that $\hat{N}$ is light so that at low energy the visible sector consists of the NMSSM \cite{Ellwanger:2009dp}, i.e. the MSSM extended by a gauge singlet. 
As freeze-in is due to a higher dimensional operator, the relic density is ultraviolet sensitive \cite{Elahi:2014fsa} and will be determined by the reheating temperature at the end of inflation, $T_{\rm rh}$. The standard Boltzmann equation applies to the evolution of the number density of states in the hidden sector:
\begin{equation} \label{eq:boltzman}
	\frac{d n_{\rm HS}}{d t} + 3 H n_{\rm HS} \simeq \frac{T}{512 \pi^5} \int_0^{\infty} d s\ |\mathcal{M}|^2 \sqrt{s} K_1 (\sqrt{s}/T)\,,
\end{equation}
where $T$ is the temperature in the visible sector, $H$ the Hubble parameter and, in our model, the amplitude reads 
\begin{equation}
    |\mathcal{M}|^2 = N_c \frac{35}{4} \frac{s^2}{M_N^4}\,.
\end{equation}
An approximate solution of the Boltzmann equation \eqref{eq:boltzman} yields the following comoving number density
\begin{equation}
Y_{\rm HS} = \frac{n_{\rm HS}}{s} \simeq \frac{1575\ M_P\ T_{\rm{rh}}^3 N_c}{16 \pi^7 \ 1.66 g_*^s\ \sqrt{g_*^{\rho}}\ M_N^4}\,.
\end{equation}
This quantity depends dominantly on the physics at high scales, hence it can be computed directly in the SYM theory, without knowledge of the dynamics at low energies. Assuming that this number density is directly converted into a number density of DM candidates, we estimate
\begin{equation} \label{eq:DMrelic}
	\Omega_{\rm{DM}} h^2 \simeq 0.125 \times 10^{23}\ N_c\ \frac{T_{\rm{rh}}^3\  m_{\rm DM}}{M_N^4}\,.
\end{equation}
After re-population by freeze-in, the dark sector undergoes a non-trivial thermal history, characterised by thermalisation via self-interactions and the confinement phase transition \cite{Garani:2021zrr}. We checked that a more accurate treatment of these effects leads to results similar to our naive estimate.  
By matching Eq.~\eqref{eq:DMrelic} to the measured value, we can relate the required DM mass to the reheating temperature and mediation scale $M_N$, as shown in Fig.~\ref{fig:relic}. The region in lilac is excluded as the required DM mass is larger than $T_{\rm rh}$. In the grey region, the dark sector is produced thermally in the early Universe instead of via freeze-in, corresponding to the model-independent bound $m_{\rm DM} \gtrsim 0.4$~keV \cite{Elahi:2014fsa}. Finally, we highlighted by a red line the region where $m_{\rm DM} = 1$~MeV, as below this line the DM is too light to allow effective structure formation \cite{Boehm:2003xr,Hooper:2007tu}. This leaves an allowed band in the parameter space, with DM masses between the MeV and the reheating scale.


At low energy, where the DM mass is dynamically generated, DM interactions are described by an effective field theory where the SYM gluons and gluinos are confined into massive bound states. The original construction by Veneziano and Yankeliowicz (VY) only included one superfield $S$, corresponding to the supersymmetric gluino-ball states \cite{Veneziano:1982ah}. Later the model was generalised to include glue-ball states in terms of a chiral superfield $\chi$ (gVY). The Lagrangian for the model reads \cite{Merlatti_2004,Feo_2004}:
\begin{multline} \label{eq:gVY}
    \mathcal{L}_{\rm{gVY}}  = \frac{9 N_c^2}{\alpha} \int d^4 \theta\  (S^{\dagger} S)^{\frac{1}{3}} \left( 1  + \gamma\ \chi \chi^{\dagger} \right) \\  +  \frac{2 N_c}{3} \int d^2 \theta\ 
     \left[ S \left(  \log \left( \frac{S}{\Lambda^3} \right)^{N_c} 
     -N_c \right)\right. \\
     \left. \phantom{\left(\frac{2}{3}\right)^{N_c}} -N_c S\ \log \left( -e \frac{\chi}{N_c} \log  \chi^{N_c} \right)  \right] +\rm{h.c}\,.
\end{multline}
Besides the number of colours $N_c$ and the confinement scale $\Lambda$, the interactions depend on two parameters: $\alpha$ and $\gamma$ ($e$ is the Euler's number). The former, mainly controls the overall strength of the couplings. The latter, instead, crucially controls the spectrum of the theory, which can be obtained from the action in Eq.~\eqref{eq:gVY} \cite{Feo_2004} (see the supplementary material for further details). After diagonalisation, the two mass eigenvalues can be written as
\begin{equation}
    m_{L/H} = \alpha \Lambda\ \mu_{L/H} (\gamma)\,,
\end{equation}
where $\alpha\Lambda$ sets the scale, while the mass ratio and the mixing only depends on the parameter $\gamma$. For $\gamma \to 0$, the glue-ball states decouple and the light state consists of pure gluino-balls (as in the original VY model) with mass $m_L \equiv m_S = 2/3 \ \alpha \Lambda$, while for larger $\gamma$ a mixing is always in place.
In principle, $\gamma$ is not a free parameter as it is fully determined by the dynamics of the SYM interactions. 
Large-$N_c$ arguments support that the gluino-balls should be lightest \cite{Feo_2004}, pointing towards small $\gamma \lesssim 1$. On the other hand, perturbative arguments suggest lightest glue-balls \cite{Farrar:1997fn,Farrar:1998rm}. Lattice results are available for $N_c=2$ \cite{Bergner:2015adz,Ali:2019gzj} and $N_c=3$ \cite{Ali:2016zke}, however unable to resolve the question yet \cite{Bergner:2022snd}.  Note that \cancel{SUSY} effects can be included in the form of a gaugino mass \cite{MASIERO1985593,Farrar_1999}, however as $m_{\rm HS} \sim$~eV, this effect can be neglected for our work.

\begin{figure}[htbp]
\centering
\includegraphics[width=8.5cm]{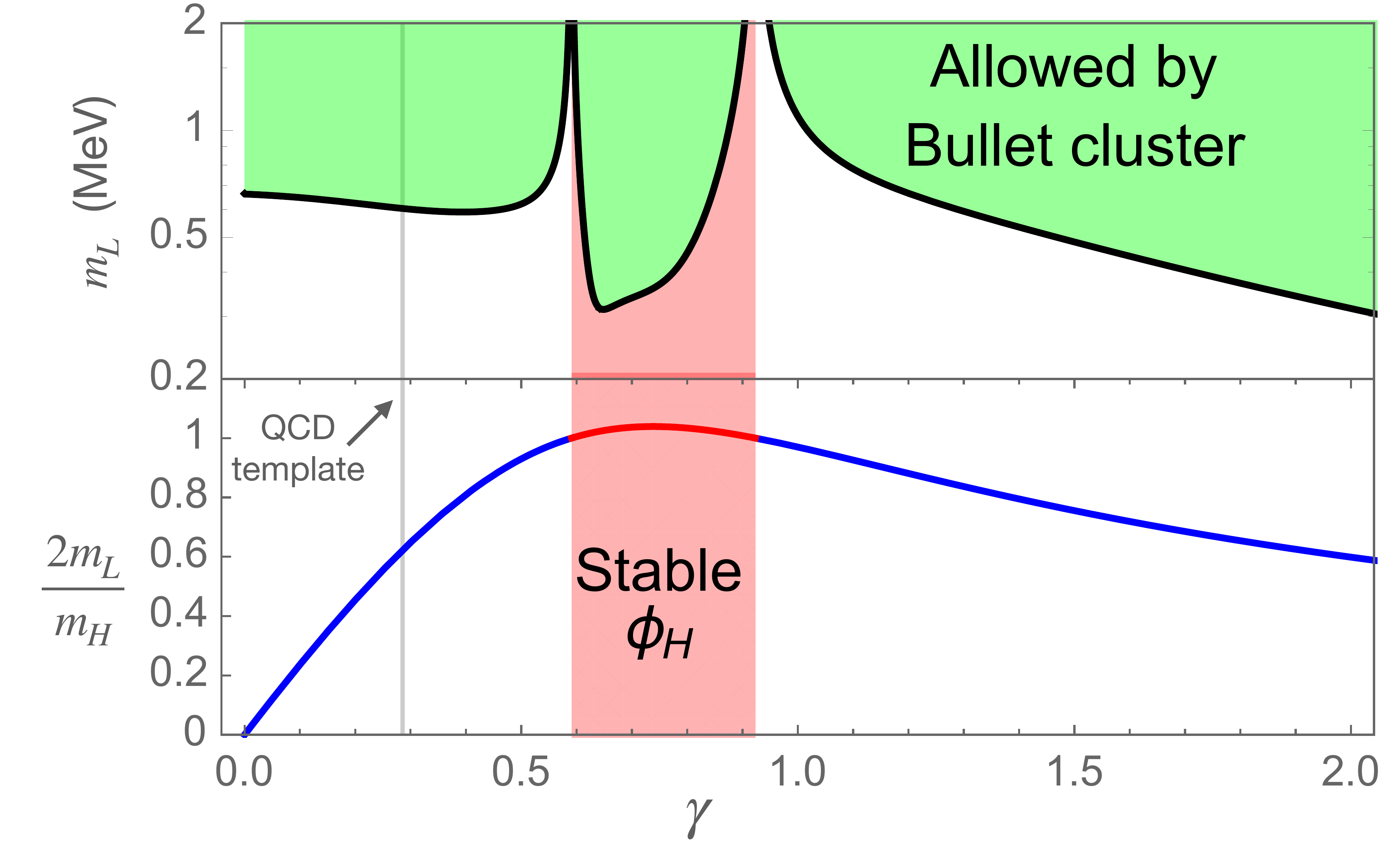}
\caption{\label{fig:bullet} Lower limit on $m_{\rm DM} = m_L$ stemming from the Bullet cluster bound on the self-scattering cross section, as a function of $\gamma$ and for $\alpha=1$ and $N_c=3$ (top panel). The bound is compared to the ratio of the mass eigenvalues (bottom panel). The red band corresponds to a two-component DM case, where the bullet bound should be considered as a conservative estimate. The vertical line marks the QCD-inspired benchmark value of $\gamma$.}
\end{figure}

The ratio of the two masses only depends on $\gamma$, as shown in the bottom panel of Fig.~\ref{fig:bullet}. Only within the range $0.59 \lesssim \gamma \lesssim 0.92$ the heavy state cannot decay into two light ones, hence the model would predict a two-component DM model. This region is highlighted in red in the figure. For all other values of $\gamma$, the DM is constituted purely of the lightest states. One-flavour QCD can be used to estimate the mass ratio, as suggested in Ref.~\cite{Feo_2004}, giving the value of $\gamma_{\rm QCD} \sim 0.29$, which we could consider as a benchmark value for $\gamma$.

The supersymmetric Lagrangian in Eq.~\eqref{eq:gVY} contains self-interactions which are mainly controlled by $\alpha$ and $\gamma$. The DM self scattering is bound by the Bullet cluster observation, providing a generic  upper limit $\frac{\sigma_{\rm{DM}}}{m_{\rm{DM}}} \leq 2\ \rm{cm}^2\cdot\rm{g}^{-1}$ \cite{Robertson_2016}, where $\sigma_{\rm{DM}}$ contains all $2\to 2$ self-scattering at low velocity. In our model, it is sufficient to consider the self-scattering of the lightest mass eigenstate, $\varphi_L$. Hence, we computed all the scattering cross sections of the scalar components, $\sigma (\varphi_L \varphi_L^\dagger)$, $\sigma (\varphi_L \varphi_L)$ and $\sigma (\varphi_L^\dagger \varphi_L^\dagger)$ to all allowed final states. Assuming that the DM halo has an equal distribution of all DM components, $\sigma_{\rm DM}$ is replaced by the average of the cross sections, as detailed in the supplementary material. The final cross section has the form
\begin{equation}
    \sigma_{\rm DM} = \frac{\alpha^6}{N_c^4} \frac{|\tilde{\mathcal{A}} (\gamma)|^2}{128\pi\ m_L^2}\,,
\end{equation}
where the effective amplitude in the numerator is a pure function of $\gamma$, depending on the non-trivial mixing between the two states in the gVY model. Hence, the Bullet cluster observation imposes a lower limit on $m_L$, which is shown in the top panel of Fig.~\ref{fig:bullet} for $\alpha = 1$ and $N_c = 3$, as a function of $\gamma$. Due to the scaling of the cross section, the limit for other parameters can be obtained by keeping $N_c^{4/3} m_L/\alpha^{2}$ constant.
The features appearing at the edge of the red band are due to the resonant contribution of the heavy state, when $m_H \sim 2 m_L$, enhancing the cross section. 
SYM also suffers from the presence of domain walls due to the $N_c$ minima \cite{Merlatti:2005sd}: their surface tension, proportional to $\sigma \propto N_c^2 \Lambda^3 \sim (N_c^2/\alpha^3)\ m_{\rm DM}^3$, is bound to be below the MeV scale \cite{Zeldovich:1974uw}. 
Hence, a non-trivial limit on the DM mass is imposed by the sizeable self-interactions of the lightest DM state and by domain walls, which further reduce the available parameter space in Fig.~\ref{fig:relic} and limits the DM mass to be around the MeV.
Heavier resonances could also contribute to the self-scattering, however this effect crucially depends on the mass spectrum. One-flavour QCD on the lattice suggests the presence of relatively light spin-1 resonances \cite{DellaMorte:2023ylq}. SYM has also been considered as a model for inflation \cite{Channuie:2012bv}, however requiring a too large composite scale to also provide a good DM candidate.

\vspace{0.3cm}

The general proposal depicted in Fig.~\ref{Model} has been illustrated with a simple $\mathcal{N}=1$ SYM theory. This scenario can support a variety of other hidden supersymmetric sectors to generate a DM candidate, hence providing a phenomenological application to many different theories. For instance, by extending the supersymmetry to the maximal type in four dimensions, $\mathcal{N}=4$ \cite{Brink:1976bc}, the theory can feature a superconformal phase \cite{Caron-Huot:2011zgw}. Furthermore, this theory is believed to be solvable and a lot of theoretical studies are ongoing \cite{Arkani-Hamed:2022rwr}, including applications of the AdS/CFT duality \cite{Maldacena:1997re,DHoker:2002nbb}. The case of conformal thermal DM has been studied in general in Ref.~\cite{Hong:2022gzo}, finding a good description of the DM relic density for masses around the MeV. Another interesting class of theories yields calculable low energy interactions by using duality properties: the prime examples are $\mathcal{N}=2$ Seiberg-Witten theories \cite{Seiberg:1994rs,Seiberg:1994aj}. Furthermore, SQCD has been considered as a source of self-interacting DM \cite{Csaki:2022xmu} to solve puzzles in structure formation. All these theories endorse the dark sector with clear predictive power, which has not been fully exploited for the DM phenomenology, yet.

\vspace{0.3cm}

In conclusion, our general proposal opens up a new avenue for the constructions of models that may explain the presence of dark matter in the universe. Supersymmetry endorses the dark sector with calculability and predictive power, also motivating a detailed study of supersymmetric theories on the lattice. In the supersymmetric dark matter scenario, both spins are always present with degenerate masses. This could lead to observable features in cosmology and astrophysics. For instance, Bose-Einstein condensation may occur in high density regions due to the presence of bosons degenerate with the fermionic partners \cite{Abt:2014nwa,blaizot:hal-01669562}. Bose-Einstein condensate have also been considered as the source for DM itself \cite{Boehmer:2007um}. Furthermore, if the sDM candidates can be effectively captured by neutron stars, its presence could be revealed in neutron star merger events \cite{Ellis:2017jgp}. The spin of the DM particles may give characteristic modifications to the neutron star elasticity, hence giving testable effects on the multi-messenger signals coming from the merger \cite{particles5030024}.
In this letter we have explored a simple coupling between the visible and hidden sectors, however other possibilities remain to be explored, providing potential links between the supersymmetric dark sector with inflation and baryogenesis.

\vspace{0.2cm}

\section*{Acknowledgements}

We acknowledge inspiring discussions with S.Hohenegger and F.Sannino.

\bibliography{bib}

\newpage
\widetext

\appendix

\setcounter{equation}{0}
\renewcommand{\theequation}{S.\arabic{equation}}
\setcounter{figure}{0}
\renewcommand{\thefigure}{S.\arabic{figure}}

\section*{Supplementary material}

Below the confinement scale $\Lambda$, the $\mathcal{N}=1$ SYM model is described by a generalised Veneziano-Yankielowicz (gVY) supersymmetric Lagrangian \cite{Merlatti_2004,Feo_2004}. It contains the following K\"ahler potential:
\begin{equation}
    K (\tilde{S},\tilde{S}^\dagger,\chi,\chi^\dagger) = \frac{9 N_c^2 \Lambda^2}{\alpha} \; (\tilde{S}^\dagger \tilde{S})^{\frac{1}{3}} (1+\gamma\ \chi^\dagger \chi)\,,
\end{equation}
and a superpotential
\begin{equation}
    W(\tilde{S}, \chi) = \frac{2 N_c \Lambda^3}{3}\; \tilde{S} \left[ \log \tilde{S}^{N_c} - N_c - N_c \log \left( - e \frac{\chi}{N_c} \log \chi^{N_c}\right) \right]\,,
\end{equation}
where we have defined a dimension-less gluino-ball field $\tilde{S} = S/\Lambda^3$. The superpotential implies the presence of a supersymmetric vacuum, characterised by
\begin{equation}
    \frac{\partial W}{\partial \tilde{S}} = 0\,, \quad \frac{\partial W}{\partial \chi} = 0\,,
\end{equation}
whose solutions describe the $N_c$ vacua of SYM theories \cite{Merlatti_2004}:
\begin{equation}
    \chi_0 = \frac{1}{e} \exp{\left(-2\pi i \frac{k}{N_c}\right)}\,, \quad \tilde{S}_0 = \exp{\left(-2\pi i \frac{k}{N_c}\right)}\,, \quad \text{where}\;\; k = 0,\dots N_c-1\,.
\end{equation}
In the following, we will consider the vacuum with $k=0$.

\begin{figure}[htbp]
\centering
\includegraphics[width=8cm]{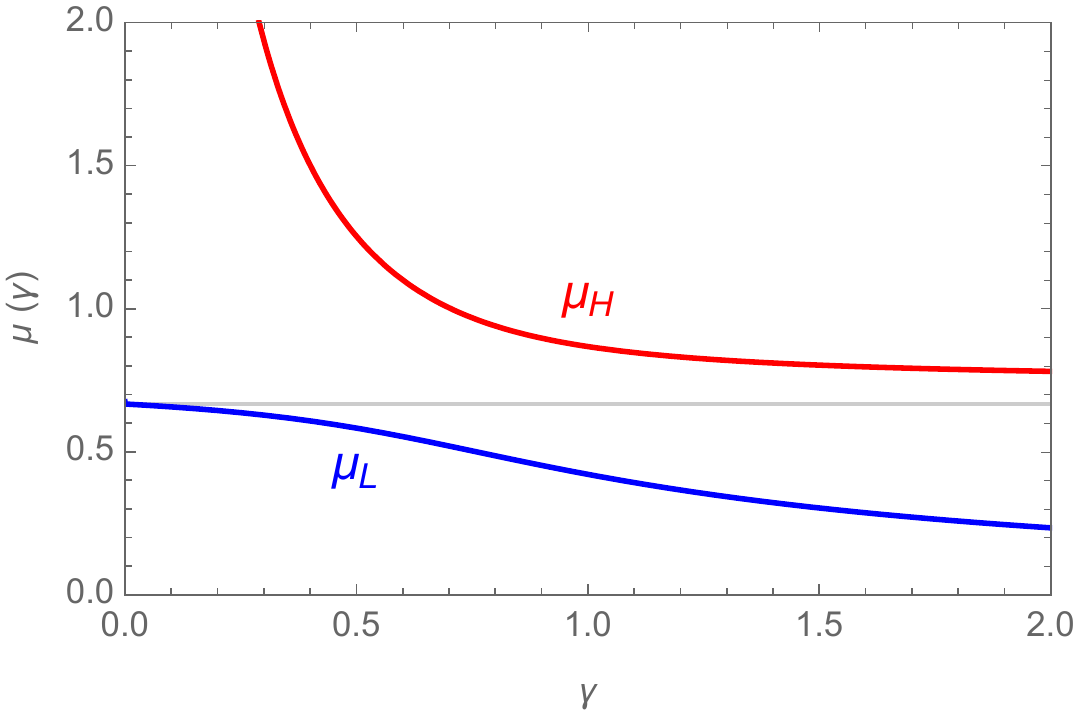}
\caption{\label{fig:masses} function of $\gamma$ defining the mass eigenvalues of the two states in the gVY model. The horizontal line corresponds to the pure gluino-ball mass, achieved for $\gamma = 0$, as $\mu_L(0) = 2/3$.}
\end{figure}

For our purposes, it suffices to study the spectrum and interactions of the scalar components, as the properties of the fermionic partners are tied by supersymmetry. Expanding the superfield Lagrangian, one obtains the standard formula
\begin{equation}
    \mathcal{L}_{\rm scalars} = \partial_\mu \varphi^l g_{lm} \partial^\mu \varphi^{\dagger,m} - \frac{\partial W}{\partial \varphi^l} g^{lm} \frac{\partial W^\dagger}{\partial \varphi^{\dagger,m}}\,,
\end{equation}
where $m,l=\tilde{S},\chi$ labels the two superfields and
\begin{equation}
    g_{lm} = \frac{\partial^2 K}{\partial \varphi^l \partial \varphi^{\dagger,m}}\,, \quad g^{l,m} = \left( g_{l,m}^{-1} \right)^T\,,
\end{equation}
are the K\"ahler metric and its inverse, respectively. To obtain the mass eigenstates, it suffices to expand the above Lagrangian, diagonalise and normalise the kinetic term, and finally diagonalise the resulting mass term, as described in Ref.~\cite{Merlatti_2004}. Finally, one obtains the following masses
\begin{equation}
    m_{L,H} = \alpha \Lambda\ \mu_{L,H} (\gamma)\,,
\end{equation}
where the dimensionless functions $\mu_{L,H}$ only depend on $\gamma$. The numerical values are shown in Fig.~\ref{fig:masses}.
In the limit $\gamma \to 0$, the glueball field $\chi$ decouples (it becomes an auxiliary field with no kinetic term), and the only remaining state (gluino-ball) has mass
\begin{equation}
    m_S = \frac{2}{3} \alpha \Lambda\,.
\end{equation}

\begin{figure}[htbp]
\centering
\includegraphics[width=8cm]{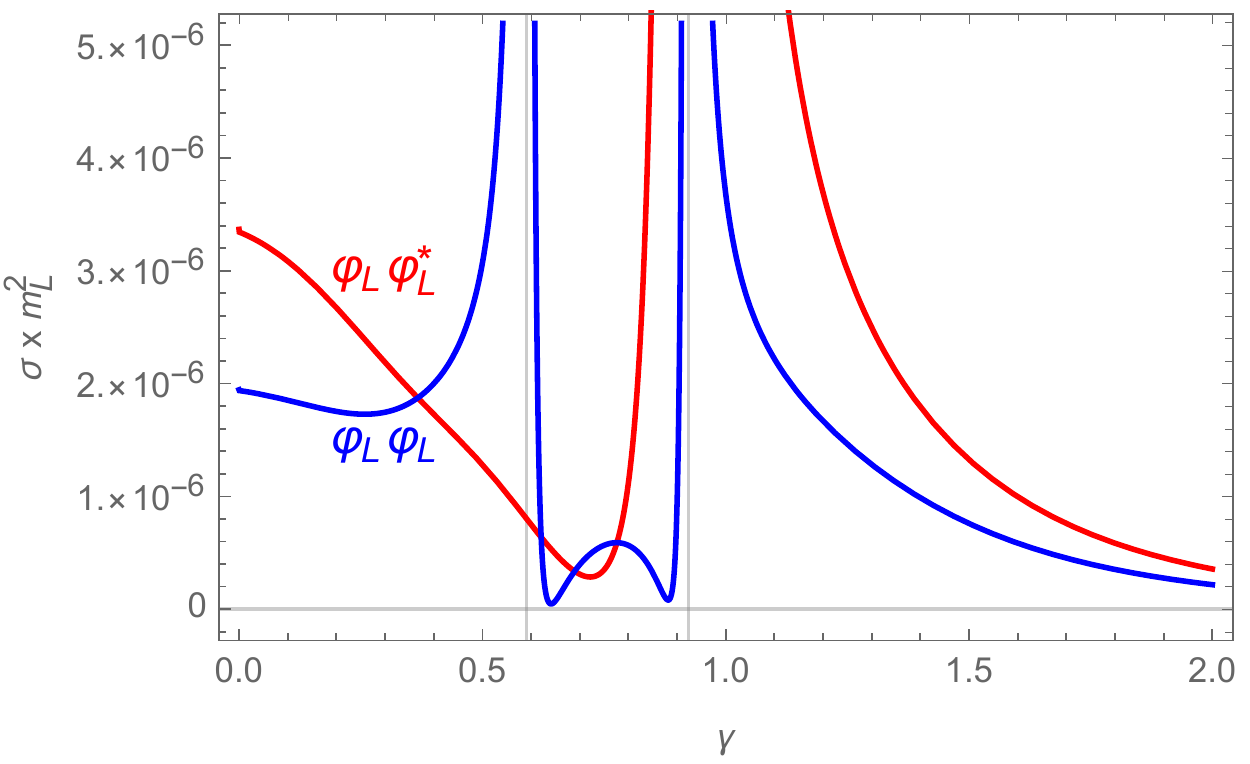}
\caption{\label{fig:cxsec} Cross sections in units of the mass for the two relevant processes, as a function of $\gamma$. The vertical lines highlight the values of $\gamma$ for which $m_H = 2 m_L$.} 
\end{figure}

To estimate the self-interactions of the DM candidate, we focus on the self-scattering of the scalar mode of the lightest mass eigenstate. Here, we will assume that the DM halo is equally populated by the various components, including their antiparticles, and that the cross sections involving fermions are related to those with pure scalars by supersymmetry. Hence, the effective DM cross section will be an average of all the possible scalar self-scattering:
\begin{equation}
    \sigma_{\rm DM} = \frac{2 \sigma (\varphi_L \varphi_L^\dagger) + \sigma (\varphi_L \varphi_L) + \sigma (\varphi_L^\dagger \varphi_L^\dagger)}{4} = \frac{\sigma (\varphi_L \varphi_L^\dagger) + \sigma (\varphi_L \varphi_L)}{2}\,,
\end{equation}
where we include all possible final states, and we use the identity $\sigma (\varphi_L \varphi_L) = \sigma (\varphi_L^\dagger \varphi_L^\dagger)$.
\begin{figure}[htbp]
\centering
\includegraphics[width=16cm]{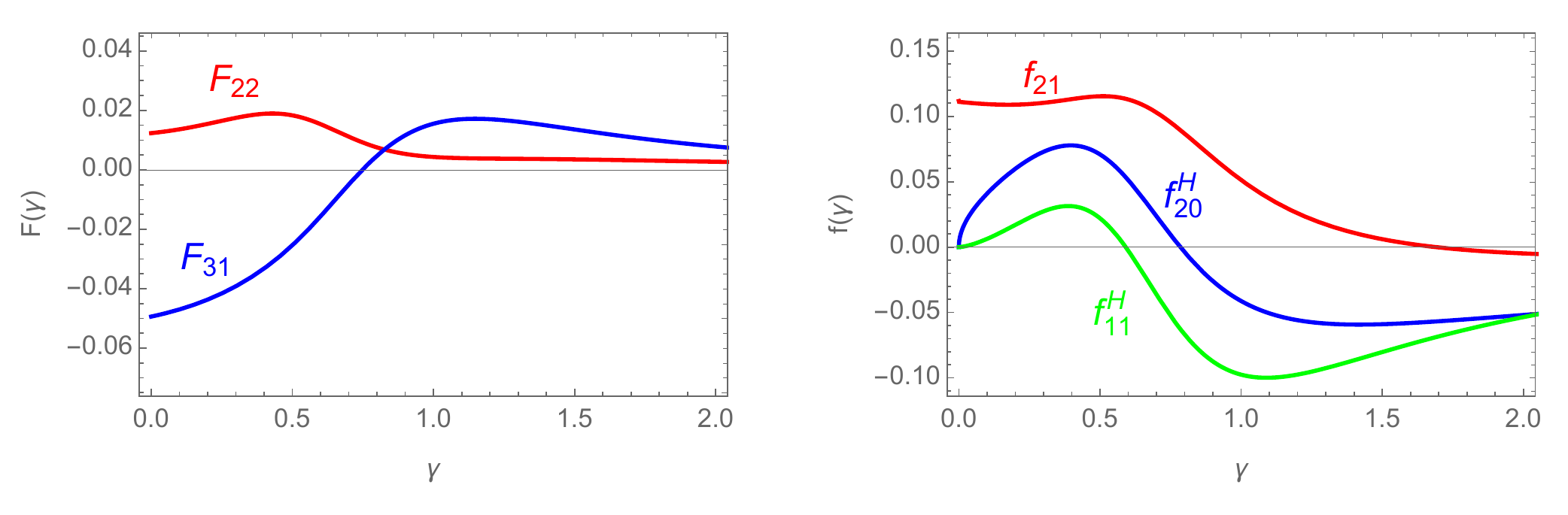}
\caption{\label{fig:couplings} Functions of $\gamma$ defining the quartic (left) and trilinear (right) couplings of the scalar components, relevant for the self-scattering of the lightest state.}
\end{figure}
The potential generates both quartic and trilinear interactions. The ones relevant for our purpose can be parameterised as
\begin{equation}
    \mathcal{L} \supset C_{31}\ (\varphi_L^3 \varphi_L^\dagger + \text{h.c.}) + C_{22}\ \varphi_L^2 (\varphi_L^\dagger)^2 + m_L \left( c_{21}\ \varphi_L^2 \varphi_L^\dagger + \text{h.c.} \right) + m_H \left(c_{20}^H\ \varphi_L^2 \varphi_H^\dagger + c_{11}^H\ \varphi_L \varphi_L^\dagger \varphi_H^\dagger + \text{h.c.} \right)\,.
\end{equation}
An explicit calculation shows that the couplings $C_x$ and $c_x^{(H)}$ can be written as
\begin{equation}
    C_x = \frac{\alpha^3}{N_c^2}\ F_x(\gamma)\,, \quad c_x = \sqrt{\frac{\alpha^3}{N_c^2}}\ f_x (\gamma)\,, \quad c_x^H = \sqrt{\frac{\alpha^3}{N_c^2}}\ f_x^H (\gamma)
\end{equation}
where $F$, $f$ and $f^H$ are functions of $\gamma$ only. The values of these functions are shown numerically in Fig.~\ref{fig:couplings}.
The amplitudes for the $\varphi_L \varphi_L^\dagger$ scattering processes at zero velocity are given by
\begin{eqnarray}
    i \mathcal{A} (\varphi_L \varphi_L^\dagger \to \varphi_L \varphi_L^\dagger) &=& \frac{\alpha^3}{N_c^2} \left[ 4 F_{22} + \frac{20}{3} f_{21}^2 + 4 (f_{20}^H)^2 + 4 (f_{11}^H)^2 \left(1 - \frac{1}{4\zeta-1} \right)\right]\,, \\
    i \mathcal{A} (\varphi_L \varphi_L^\dagger \to \varphi_L \varphi_L) &=& \frac{\alpha^3}{N_c^2} \left[ 6 F_{31} + \frac{20}{3} f_{21}^2 + 2 f_{20}^H f_{11}^H \left( 2 - \frac{1}{4\zeta-1} \right)\right]\,, \\
    i \mathcal{A} (\varphi_L \varphi_L^\dagger \to \varphi_L^\dagger \varphi_L^\dagger) &=& i \mathcal{A} (\varphi_L \varphi_L^\dagger \to \varphi_L \varphi_L)\,;
\end{eqnarray}
where $\zeta = m_L^2/m_H^2$, being a pure function of $\gamma$.
For the $\varphi_L \varphi_L$ scattering, we have
\begin{eqnarray}
    i \mathcal{A} (\varphi_L \varphi_L \to \varphi_L \varphi_L) &=& \frac{\alpha^3}{N_c^2} \left[ 4 F_{22} + \frac{20}{3} f_{21}^2 - 4 (f_{20}^H)^2 \left(\frac{1}{4\zeta-1} \right) + 8 (f_{11}^H)^2 \right]\,, \\
    i \mathcal{A} (\varphi_L \varphi_L \to \varphi_L \varphi_L^\dagger) &=& i \mathcal{A} (\varphi_L \varphi_L^\dagger \to \varphi_L \varphi_L)\,.
\end{eqnarray}
Finally,
\begin{equation}
    \sigma (\varphi_L \varphi_L^\dagger) = \sum_f \frac{|\mathcal{A} (\varphi_L \varphi_L^\dagger \to f)|^2}{128 \pi\ m_L^2}\,, \quad \sigma (\varphi_L \varphi_L) = \sum_f \frac{|\mathcal{A} (\varphi_L \varphi_L \to f)|^2}{128 \pi\ m_L^2}\,,
\end{equation}
where the sum runs over all allowed final states.
As mentioned above, the average of these cross sections is used to estimate the bound on $m_L$ from the Bullet cluster. The two cross sections are plotted in Fig.~\ref{fig:cxsec} as a function of $\gamma$ and in units of the mass. While $\sigma (\varphi_L \varphi_L)$ clearly shows the presence of the two resonances in the $s$-channel, where $m_H = 2 m_L$, one resonance is missing for $\sigma (\varphi_L \varphi_L^\dagger)$. This is due to the fact that, in the latter, the $s$-channel is proportional to the coupling $c_{11}^H$, which vanished at the resonance, as shown in Fig.~\ref{fig:couplings}.

\end{document}